# ASTROPHYSICAL PARADOXES, long version

DRAGOLJUB A. CUCIC

Regional centre for talents “Mihajlo Pupin”, Pancevo, Serbia, rctpupin@panet.rs

ABSTRACT

Astrophysical paradoxes are the paradoxes of physics. The main motivation of a formulated paradox is clearly recognized in the scientific environment because the phenomenon of a paradox itself has become interesting. There is an explanation of how and why the phenomenon of paradox started to exist, as there is an explanation for the existence of any phenomenon. A paradox has its structure, which defines the functional aim of creating paradoxes. According to the structure there are different types of paradoxes in astrophysics and some of them are going to be classified and analyzed here. Astrophysical paradoxes have mostly been solved or else there are theoretical premises for their solution. Their structure is recognizable in two distinct ways that lead to the solution through changing the paradigm or through a hierarchical sequence ending in the solution.

## Introduction

Certain problems in physics, biology, astrophysics, or other sciences have often been in some contexts called paradoxes. Instead of the term paradox modern physicists sometimes use the term *puzzle*, although that term is more suitable for a problem task that needs to be solved. In its structure a paradox contains a problem which is probably the cause of this synonym identification. A paradox is not only a problem that needs to be solved; it contains a contraintuitive element at odds with the existent explanation, which is the basis of its incongruity.

It is not the intention of this work to solve paradoxes. This would require knowledge far more complex than it is possible for one person to attain. Since the given solutions are by different authors, this work will sketch a “matrix” that will make paradoxes recognizable. This work will classify paradoxes and formally expound their solutions, should any exist. There is no clearly formulated text synthesising paradoxes in astrophysics by any classification factors. The classification of paradoxes that will be used here is the one formulated in my work "The types of paradox in physics". Paradoxes in astrophysics are essentially physical paradoxes[1] which is the reason why the physical classification is used here as well.

This text defines a fresh view on paradoxes in astrophysics and as such is subject to criticism that may lead to improvement through modification or that could refute the very argument through rational interpretation. The aim was to present several paradoxes in astrophysics and establish the existence of regularity that may be present in the formation of paradoxes in astrophysics and, finally, to set recognizable common factors for these paradoxes.

## Astrophysical paradoxes

A serious study of the phenomenon of paradox began at the end of the last century, even though the phenomenon had been present as early as in the works of ancient Greek authors. Many definitions of paradox can be found in the literature. Richard Mark Sainsbury, one of the acclaimed quoted authors who work on the phenomenon of paradox, called it an ***unacceptable conclusion resulting from an acceptable model of inference and from acceptable initial presuppositions***. A paradox is a form of contraintuitive perspective. The nature of perspective taking is manifold: theoretical, experimental, visual (observational), semantic. A paradox is a rightly established dilemma, the rightness of which is based on valid previous knowledge.

[1] Facts are obtained by the analysis of electromagnetic waves and subatomic particles coming from space.

According to the manner of establishing a paradox, and the reasons why a certain phenomenon in physics is called paradoxical, the following paradoxes in astrophysics (physics) are differentiated:

1.***Pseudo paradox*****,** (Not a real paradox. A precise analysis can establish that no actual physical incongruity exists; that a "so-called incongruity" comes from superficial examination.)

2.***Paradox of idealization,*** (Formed when a physical process is idealized and when the likelihood of the physical event is extremely small.)

3.***Hierarchical paradox*****,** (This type of paradox is characterized by the absence of explanation why a change of principle occurs in different, hierarchically differentiated[2], physical states.)

4.***Transitions paradox,*** (Formed in the process of solving a formulated idea. It is a problem step in the explanation of a theoretic or physically possible phenomenon.)

5.***Paradox of assumption*****,** (The paradox in which the initial supposition in the process of explaining a physical phenomenon is inaccurate within the framework of the same paradigm. Deductive analysis is based upon the assumption leading to the conclusion contradictory to the actual condition of the physical system.)

6.***Paradox of paradigm*****,** (This type of paradox exists within the framework of a paradigm. When the paradigm is changed the paradox is gone.)

[2] Ed. transition from a micro to macro system.

Astrophysical paradoxes cover a wide spectrum of physical branches that are used to formulate and solve astronomical phenomena. Hereafter the following paradoxes will be introduced and tested:

1. GZK paradox, (Cosmic ray paradox)
2. Seeliger's paradox,
3. Olbers' paradox (Photometric paradox),
4. Wheeler's paradox of black hole entropy,
5. Black hole information paradox (Hawking's paradox),
6. Eddington paradox,
7. Faint young Sun paradox,
8. The Heat Death paradox (Clausius paradox).

The common denominator for all these paradoxes is that they all deal with astronomical phenomena that were at one time, or still are, the starting point of an integral approach to solving a problem based on different physical theories.

### 1. GZK paradox (Cosmic ray paradox)

*There is a computed energy upper limit for the measurement of radiation originating from distant objects in the Universe. When cosmic ray energy is above this limit there is no interaction between EM radiation originating from distant objects and photons of the cosmic microwave background radiation. The paradox is that there is evidence of cosmic rays originating from distant sources with energy above the established limit.*

*GZK paradox* is based upon a predefined *GZK cutoff that* was independently calculated in 1966 by an English scientist Kenneth Greisen (1918-2007)[3], from Cornell University, and at the same time the Russians Vadim A. Kuzmin and Georgiy T. Zatsepin[4], whose initials were used to name this paradox. They theoretically determined the threshold of cosmic radiation energy from distant sources for interacting with photons of the cosmic microwave background radiation.

The computation of the interaction threshold is based on the special theory of relativity and particle physics. Lee Smolin emphasized that the *GZK paradox* prediction was:

> "*… the first test of special relativity approaching the Planck scale, the scale at which we might see the effects of a quantum theory of gravity.*"[5]

The paradox is based upon the limit for interaction between cosmic rays and photons of the cosmic microwave background radiation. Cosmic rays with energy above this limit will not interact with the photons of the cosmic microwave background radiation and no pions will be produced by the interaction. This is why the cosmic rays with energies above this threshold cannot be detected on Earth. The *GZK paradox* is that observations have shown cosmic rays with energies above the GZK cutoff (ultra-high-energy cosmic rays). There is still some vague discrepancy between the real observation results and knowledge obtained by STR and particle theory.

This paradox can otherwise be found as *Cosmic ray paradox*, and *GZK prediction*. *GZK paradox* is based upon the difference between a theoretical perspective and the results of an actual experiment. It is probably a *paradox of assumption*. If the reasons for the established difference are so great that the basic paradigmatic theoretic principles of the phenomenon explanation will have to be changed then it is a *paradox of paradigm*.

---

3 Greisen, K. (1966). End to the Cosmic-Ray Spectrum? *Phys. Rev. Lett.*, 16 (17): 748-50.
4 Zetsepin G. T., Kuzmin V.A. (1966). Upper Limit of the Spectrum of Cosmic Rays. *JTEP Letters*, **4**: 78-80.
5 Smolin, L. (2007). *The Trouble with Physics*, Mariner Books, New York. стр. 220.

This paradox originates in a real phenomenon, accessible to human sensory perception, and it is not a thought experiment but a real sensory observation.

*The solution of GZK paradox*

*GZK cutoff* is a theoretically formulated reaction limit; observation results show disagreement with the theory. There are a number of suppositions about the causes of the *GZK paradox*:

–The AGASA[6] observation results could be due to an instrument error,

–An incorrect interpretation of the AGASA observation experiment,

–Cosmic rays come from distant local sources of fairly vague origin,

–Heavier nuclei could possibly circumvent the *GZK limit*.

There is no definite solution of the *GZK paradox*, and this paradox is considered one of the current problems that the astrophysicist and physicist are yet to solve.

## *2. Seeliger's paradox*

*According to the classical, static model of the universe the stars are evenly distributed in the universe. With respect to the even distribution of stars, it can be deduced that gravitational potential is an indefinite expression. Therefore any object in the universe is subject to an indefinite gravitational potential, that is, an indefinite gravitational force originating from the other masses in the universe evenly distributed around it.*

[6] AGASA (Akeno Giant Air Shower Array) experiment refers to the observations done in Japan the results of which show energies above the GZK cutoff.

An Austrian and German astronomer, knight Hugo Hans Ritter von Seeliger (1849-1924) defined this paradox that bears his name.

The 19th century was dominated by a static classical model of the universe, founded on the basic principles: homogeneity and isotropy of space, and Euclidean time and space infinity. The very homogeneity and isotropy of the universe are the causes of the *Seeliger's paradox*. The solution of the paradox is reached through the adoption of relativistic physics that negates the classical model of the universe and changes the basic principles it was founded on; therefore this is a *paradox of paradigm*. The paradox has been solved and it is an *exparadox*.

This paradox originates in a real phenomenon, accessible to human sensory perception, and it is not a thought experiment but a *real sensory observation*.

*The solution of Seeliger's paradox*

*Seeliger's paradox* has been explained by the *Fridman's model* of the universe that rejects the assumption that matter is evenly, statically distributed in the universe. The rejection of this assumption was justified by the development of relativistic physics and Hubble's discovery of redshift, which affirmed the distancing of all the galaxies in the universe.

### *3. Olbers' paradox (Photometric paradox)*

*In accordance with the classical model of the universe, static and infinite, filled with evenly distributed stars, the brightness of the stars should evenly illuminate the entire universe. It is well know that the night sky is dark, but according to the classical model it should not be less bright than the brightness of the evenly arranged stars.*

The first who examined this contraintuitive phenomenon was a German Friedrich Johannes Kepler (1571-1630) in 1610, at the same time when an Englishman Edmund Halley (1656–1742) first put forward the dark sky phenomenon as an argument against a static universe filled with an infinite number of stars. It was a German astronomer Heinrich Olbers (1758-1840) in 1823 who formulated[7] the paradox and presented it to the scientific community. In his honour the paradox is called the *Olbers' paradox*, or the *Paradox of black night sky* as it is also known.

The paradox is a consequence of the assumption, based upon the classical model, that the universe should be entirely illuminated by the stars. We all bear witness that this is not the case. Were the universe infinite, the stars would cover the sky completely and the night sky would not be dark.

This paradox which here we classify as astrophysical, due to the complexity of astronomy, could be classified as a thermodynamical paradox, or even an electrodynamical paradox.

The paradox is based upon a theoretical perspective and it is an example of a *paradox of paradigm*. To solve it, it was necessary to amend the fundamental principles upon which the model that gave birth to the *Olbers' paradox* was made. This paradox is solved and therefore can be classified as an *exparadox*.

This paradox originates in a real phenomenon, accessible to human sensory perception, and it is not a thought experiment but a *real sensory observation*.

---

7 In 1826 he reformulated it in the present form.

*The solution of Olbers' paradox*

The discovery of redshift in 1868 by an English astronomer William Huggins (1824-1910)[8] cleared the way for a new theory. It is clear today that the universe is not static and infinite and that interstellar space is not empty. It has been confirmed that the light from distant stars brings less energy to the observer the further the galaxy or star is. That loss of energy is due to:

–Absorption by interstellar matter, which has not yet reached the state of thermodynamic equilibrium, for when this state has been reached the interstellar matter will emit as much energy as it has absorbed.

–Distancing of galaxies, since, owing to redshift, observers on Earth receive less energy, which is also the explanation of the black night sky.

It was Olbers who first proposed a solution of the paradox for the model of an infinite, stationary universe which amounted to the absorption of energy by interstellar matter. However, the solution proved inadequate since infinite universe implies that thermodynamic equilibrium should already have been reached. There have been other attempts to solve the paradox within the framework of the classical, static model like the hierarchical structure of a Swedish astronomer Carl Wilhelm Ludwig Charlier (1862–1934) in 1908 and, more recently, the Russians Kosinov, Garbaruk and Polyakov[9]. The latter do not deny the nonstationary nature of the universe; they simply offer a solution of the problem within the stationary, infinite model.

8 He discovered the Dopler effect on remote objects in the universe and named it Dopler's shift.

9 Kosinov, N. V., Garbaruk, V. I., Polyakov, D. V. Photometric Paradox and relict radiation – two sides of one phenomenon?

### *4. Wheeler's paradox of black hole entropy*

*According to the classical model and general theory of relativity nothing can leave a black hole. Any system that enters a black hole disappears inside it. Assume a complex physical system entering a black hole and a question arises: What happens with the total entropy of the black hole? What happens with the total entropy outside the black hole? Does it decrease because a part of the entropy disappears inside the black hole? If the entropy of a complex physical system disappears the second principle of thermodynamics is violated ($S \geq 0$).*

A Black hole is, according to the classical concept, an object whose gravity field is so large that nothing can escape it. [10]

*Wheeler's paradox of black hole entropy* and *Black hole information paradox* are two close related paradoxes created in the development of the theoretical explanation of the black hole phenomenon.[11] The black hole phenomenon is a product of gravitation theory, thermodynamics and quantum theory created as part of complex explanations of astrophysical phenomena during the 20th century.

*Wheeler's paradox* or *Paradox of black hole entropy* is a *theoretical paradox* that originated in the process of formulating a theory and therefore can be classified as a *transitions paradox*. *Wheeler's paradox* has been solved from a theoretical point of view in accordance with the principles of the theory and therefore is an *exparadox*.

This paradox does not originate in a real phenomenon, accessible to human sensory perception. A phenomenon is discussed that is scientifically uncertain albeit very plausible. It is speculative and originates in a phenomenon formulated in thought. This paradox is a *thought experiment*.

[10] The possibility of an astronomical body with black hole like properties was pointed out in 1783 by Englishman John Michel. The godfather of the term black hole is American physicist John Archibald Wheeler (1911-2008) who coined it in 1969.

[11] There had not been any physical evidence of a black hole and, even today, there is only a high percentage of certainty that black holes exist.

<u>*Solution of Wheeler's paradox of black hole entropy*</u>

In 1971 Stephen Hawking proved that black hole event horizon can not decrease. The reply to Wheeler's question is supplied by Jacob Bekenstein who formulated the idea that the area of the event horizon is a quantitative measure of black hole entropy. Matter enters a black hole, event horizon increases, black hole entropy increases. The total entropy of the system composed of the black hole and the outside space does not decrease.

### *5. Black hole information paradox (The Hawking's paradox)*

> *When matter enters a black hole its event horizon increases. Event horizon is a measure of entropy within a black hole. Anything that has entropy has temperature. A body with temperature emits radiation. The emission decreases the mass of the black hole. What happens with the information that disappeared in the black hole?*

This paradox originated from the exploration of gravitational effect on the basis of knowledge gained from quantum mechanics and general theory of relativity, and is founded in the idea that information cannot be destroyed.[12] The paradox could also be classified as thermodynamic since it discusses entropy and physical system information.

According to Stephen Hawking[13] the initial idea about black hole radiation started in 1967 when Werner Israel showed that Schwarzschild metrics is the only static vacuum black hole solution. In 1972 Jacob Bekenstein concluded that black holes emitted radiation[14], and a year later Stephen Hawking[15] developed the idea, which paved the way for the *black hole information paradox.*

---

[12] In view of the principle that the laws of nature are symmetrical in time, if the time direction of a certain event is reversed the necessary information will be recreated.

[13] Hawking, S.W. (2005). Information Loss in Black Holes. [hep-th/0507171]

It should be pointed out that the *black hole information paradox* resulted from a previous paradox established by Bekenstein: that black hole radiates (if a black hole is indeed black it can only absorb[16]) – which was a fantastic progress in the understanding of the black hole phenomenon that was only to be mathematically and physically shaped and confirmed.

The paradox is theoretical in nature. It is based on the theories of quantum mechanics and GTR in the attempt to explain a phenomenon the existence of which is uncertain. The assumption is that the paradox is brought about by the change of the system (black hole) physical state and therefore can be classified as a *hierarchical paradox*.

This paradox does not originate in a real phenomenon, accessible to human sensory perception. A phenomenon is discussed that is scientifically uncertain albeit very plausible. It is speculative and originates in a phenomenon formulated in thought. This paradox is a *thought experiment*.

*The Solution of the black hole information paradox*

The solution of the paradox comes from a mathematically identified and adopted "occurrence" that information is disintegrated in a black hole, based on Hawking's argument that the physical theories used[17] are temporally symmetrical. Hawking showed that the principle of reversibility in microprocesses does not apply to a black hole, because it is a phenomenon that does not allow information to escape its event horizon and in this way it constitutes a "*fundamental new source of irreversibility in nature* "[18].

---

14 Bekenstein, J. (1972). Black holes and the second law. *Lett. Nuovo. Cim.* **4**, 737-40.
15 Hawking, S. W. (1975). Particle Creation by Black Holes, *Commun. Math. Phys.* **43**.
16 A black hole behaves differently. When its mass (energy) increases it cools down.
17 GTP, thermodynamics, quantum theory

Stephen Hawking’s mathematical formalisation defined an attitude that is contrary to one of the basic quantum-mechanical principles of information indestructibility.

This paradox still has not been experimentally confirmed but in a theoretical view it has been solved and therefore is an *exparadox*. It was based on a series of suppositions that proved wrong as the theory was being formulated.

### *6. Eddington paradox*

> "*I do not see how a star which has once got into this compressed state is ever going to get out of it...*
>
> *It would seem that the star will be in an awkward predicament when its supply of subatomic energy fails*."[19]

The paradox was first made public by an English scientist sir Arthur Stanley Eddington (1882-1944) in 1926 in his book *The Internal Constitution of Stars*[20]. It was physically clarified and solved by his compatriot sir Ralph Howard Fowler (1889-1944) in his article entitled *Dense Matter*[21] and published that same *year*:

> "*The stellar material, in the white-dwarf state, will have radiated so much energy that it has less energy than the same matter in normal atoms expanded at the*

---

18 Susskind, L. (1997). Black Holes and the Information Paradox. *SCIENTIFIC AMERICAN*
19 Adopted from: Chandrasekhar, S. (1983). On Stars, their evolution and their stability. Nobel lecture.
20 Eddington, A. S. (1926). *The Internal Constitution of Stars*. Cambridge University Press.
21 Fowler, R. H. (1926). *Mon. Not. Roy. Astr. Soc.*, 87, 114.

*absolute zero of temperature. If part of it were removed from the star and the pressure taken off, what could it do*?"[22]

In aphorism form *Eddington paradox* says: "*a star will need energy to cool*."[23] Sir Ralph Fowler solved the paradox by using *Fermi's statistics* and *electron degeneracy*[24] answering at the same time the question he was asking, therefore this is an *exparadox*. *Eddington paradox* is a theoretical paradox originating in the process of solving a certain problem by applying a different mathematical mechanism. According to the adopted classification the paradox can be classified among the *hierarchical paradoxes* since it occurs during the change in the state of the matter.

This paradox does not originate in a real phenomenon, accessible to human sensory perception. A phenomenon is discussed that is far removed from human reach. It is speculative and originates in a phenomenon formulated in thought. This paradox is *a thought experiment*.

*The solution of Eddington paradox*

The solution of *Eddington paradox* according to Fowler is that electrostatic energy per unit volume of atoms is less than kinetic energy of thermal motions per unit volume of free particles in the form of a perfect gas. This causes an unequal dependence of pressure from density, which will be disrupted when the pressure is large enough (as in the case of a white dwarf). It is to be expected that electrons are degenerate when temperature and density are those that exist on white dwarfs. When densities are those that exist on white dwarfs the total kinetic energy is larger than the potential energy, which is the solution of *Eddington paradox*. As Fowler himself concluded:

[22] Adopted from: Chandrasekhar, S. (1983). On Stars, their evolution and their stability. *Nobel lecture*.
[23] Adopted from: Chandrasekhar, S. (1983). On Stars, their evolution and their stability. *Nobel lecture*.
[24] Used in this context for the first time.

"*The only difference between black-dwarf matter and a normal molecule is that the molecule can exist in a free state while the black-dwarf matter can only so exist under very high external pressure.*"[25]

***7. Faint young Sun paradox***

*At one time the volume of the Sun was smaller than it is today. The surface of the Sun was smaller and the Sun emitted less light and heat. Research has shown that, at the time, the surface of the Earth was warmer than today. How could the Sun that emitted less energy heat the Earth more*?

The first observations, from which the paradox resulted, were made by Carl Sagan (1934-1996) and George Mullen in 1972.

The observations were based upon the standard model of the Sun used to describe the development of sun-like stars. 4.5 billion years ago the Sun emitted 70% less energy than today and its volume was about 15% smaller. Under these conditions the Earth received 30% less energy from the Sun. According to the parameters valid on Earth today it should have been completely frozen. It is well known that our planet at the time in question was not frozen but warmer than today.

*The faint young Sun paradox* is a solved paradox and therefore it is an *exparadox*. The solution of the paradox is simply a validated assumption about the atmospheric conditions on Earth at the time its atmosphere and surface were formed. Sagan and Mullen knew of these conditions, which makes this paradox more like a trick question. This is why I classify it as a *pseudo paradox*.

---

[25] Adopted from: Chandrasekhar, S. (1983). On Stars, their evolution and their stability. *Nobel lecture*.

This paradox does not originate in a real phenomenon, accessible to human sensory perception. A phenomenon is discussed that took place before the existence of man. It is speculative and originates in a phenomenon formulated in thought. This paradox is *a thought experiment*.

<u>*The solution of the faint young Sun paradox*</u>

The solution of the paradox can be found in the composition of the Earth's atmosphere. Long ago the Earth's atmosphere was mainly made up of carbon-dioxide, methane and water (like that of Venus today), which created the greenhouse effect. The surface of the Earth is not now as warm since the structure of the atmosphere is significantly different, which makes the effect less evident.

### *8. Heat Death paradox (Clausius' paradox)*

*Assuming that the universe is eternal, a question arises: How is it that thermodynamic equilibrium has not already been achieved?*

*The Heat death paradox*, otherwise known as C*lausius' paradox* and *Thermodynamic paradox*, is founded on the basic assumption that each system tends to achieve thermodynamic equilibrium. This paradox is based upon the classical model of the universe in which the universe is eternal. C*lausius' paradox* is *paradox of paradigm*. It was necessary to amend the fundamental ideas about the universe, which brought about the change of the paradigm. The paradox was solved when the paradigm was changed. The paradox which is valid in the classical stationary model of the universe is not so in Fridman's nonstationary relativistic model. This is a solved paradox and therefore is an *exparadox*.

The paradox was based upon the rigid mechanical point of view of the *Second principle of thermodynamics* postulated by a German physicist Rudolf Julius Emanuel Clausius (1822-1888) according to which heat can only be transferred from a warmer to a colder object. If the universe was eternal, as claimed in the classical stationary model of the universe, it would already be cold.

This paradox originates in a real phenomenon, accessible to human sensory perception, and it is not a thought experiment but *a real sensory obs*ervation.

*The solution of the Heat Death paradox*

According to recent cosmological theories the universe is not eternal and it began some 15 billion years ago. In view of this, the solution of the *Heat death* p*aradox* is that thermodynamic equilibrium has not been achieved because not enough time has passed.

## Conclusion

In order to solve astrophysical paradoxes one needs a very extensive knowledge of physics and contemporary physical theories. As far as I know this sort of analysis has never been performed in astrophysics. Most astrophysical paradoxes have been solved and can be categorized as *exparadoxes*. It is necessary to point out that a number of the paradoxes are theoretical in nature and this theoretical nature and the fact that they are based on certain fundamental principles have to be taken into account when their solutions are discussed.

The paradoxes in astrophysics were formed in two ways: in the process of explaining astrophysical phenomena that had been observed and during the development of physical theories that assumed the existence of certain astrophysical phenomenon.

This work classifies astrophysical paradoxes according to a previously adopted classification. It can be noticed that there are no astrophysical paradoxes belonging to the type of paradox of idealization, while all other classification types are present. The majority of paradoxes are *paradoxes of paradigm* (paradoxes that originated within the framework of the classical, static model of the universe and that were solved with the adoption of Fridman's model of the dynamic universe) and *hierarchical paradoxes* (originating in the process of formalization of theories while there is no system change).

**Acknowledgements**

I wish to express my gratitude to Professor Darko Kapor for his suggestions that helped me define certain viewpoints more precisely.

| Pseudo paradox | Paradox of idealization | Hierarchical paradox | Paradox of paradigm | Paradox of assumption | Paradox of paradigm |
|---|---|---|---|---|---|
| The faint young Sun paradox | | Black hole information paradox | Wheeler's paradox | GZK paradox | Olbers' paradox |
| | | Eddington paradox | | | Seeliger's paradox |
| | | | | | Clausius' paradox |

*Table 1*. Table representation of paradoxes in astrophysics.

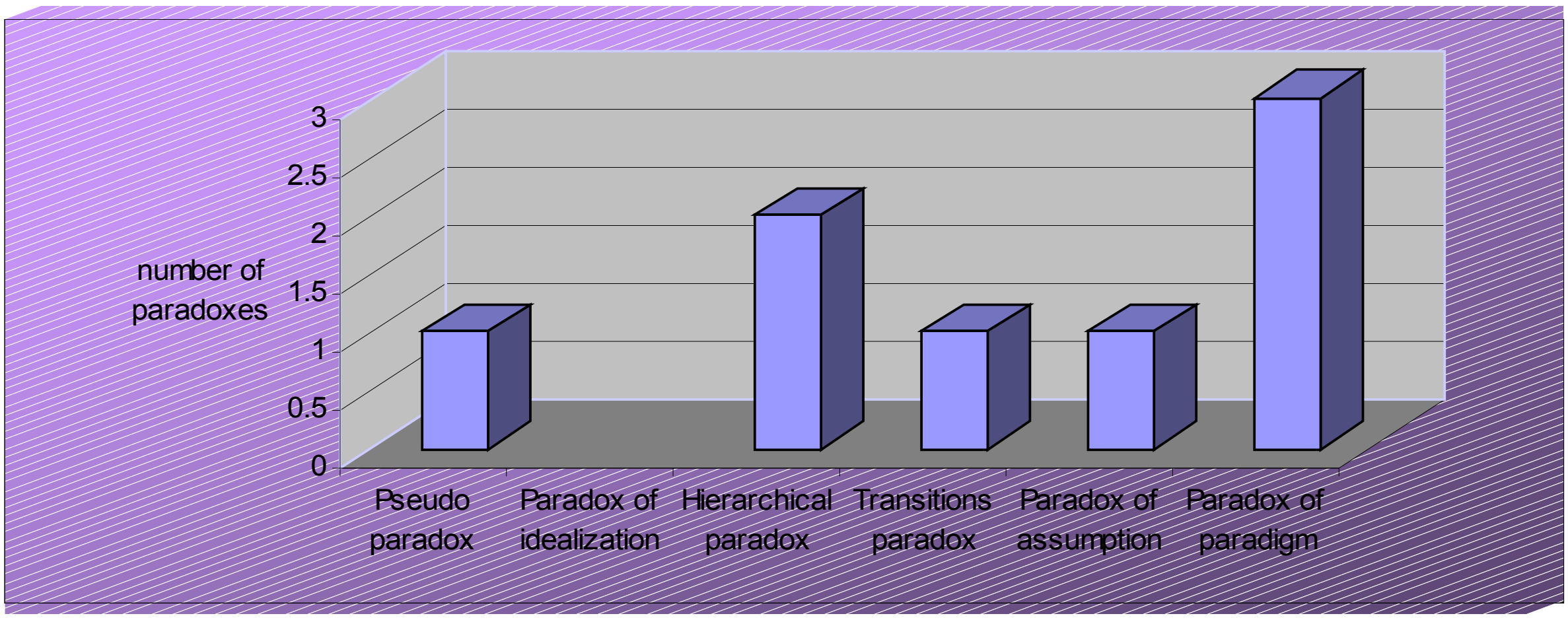

Graphic 1. Graphical representation of paradoxes in astrophysics.

| | | | | | | | |
|---|---|---|---|---|---|---|---|
| ***Exparadox*** | Clausius' paradox | *Seeliger's paradox* | Olbers' paradox | Eddington paradox | The faint young Sun paradox | Black hole information paradox | Wheeler's paradox |
| **Unsolved paradox** | GZK paradox | | | | | | |

*Table 2.* Table representation of solvability of paradoxes in astrophysics.

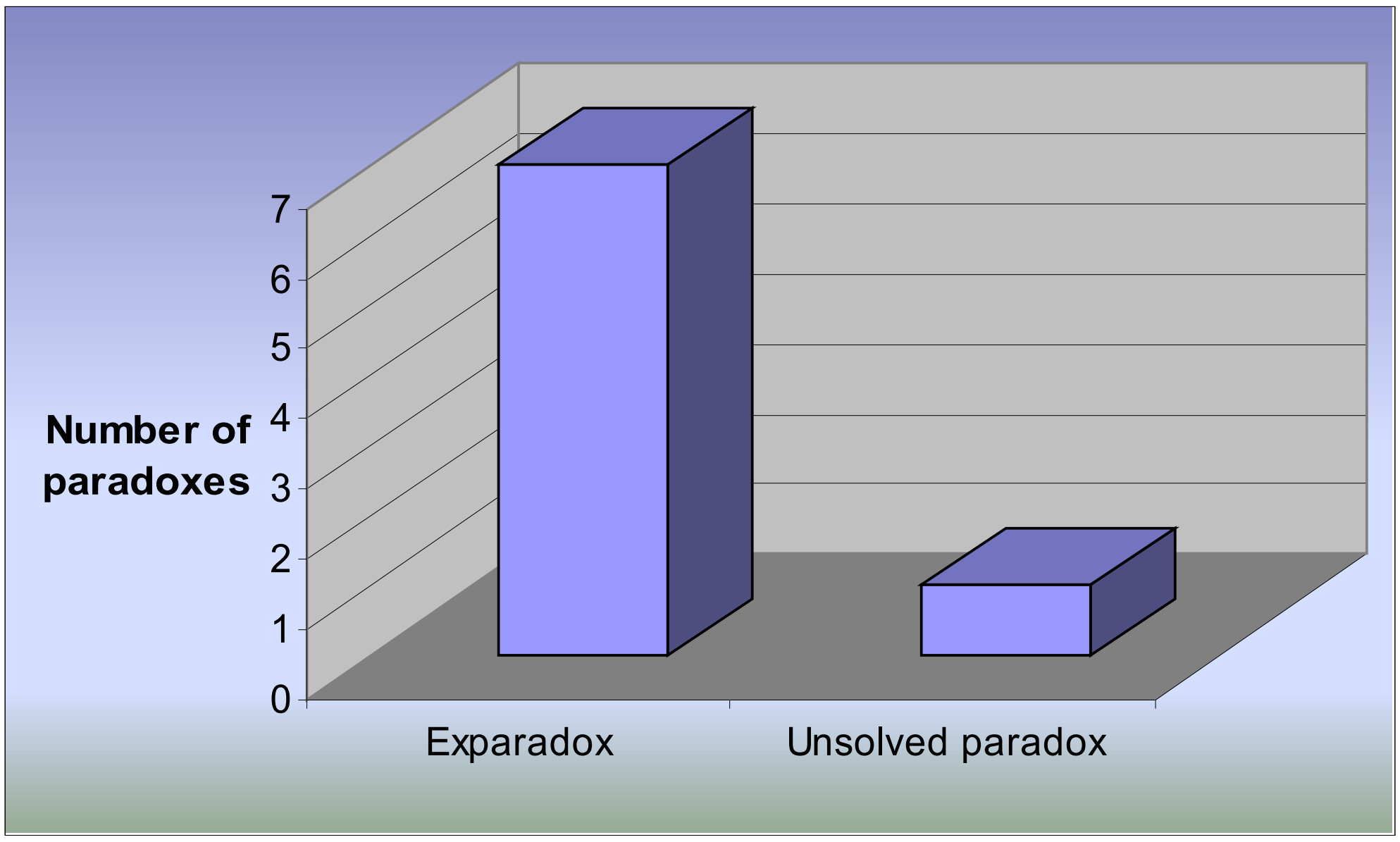

Graphic 2. Graphical representation of solvability of paradoxes in astrophysics.

| | | | | |
|---|---|---|---|---|
| **Real sensory observation** | Clausius' paradox | Seeliger's paradox | Olbers' paradox | GZK paradox |
| **Thought experiment** | Eddington paradox | Black hole information paradox | Wheeler's paradox | The faint young Sun paradox |

*Table 3.* Table representation of thougth paradoxes in astrophysics.

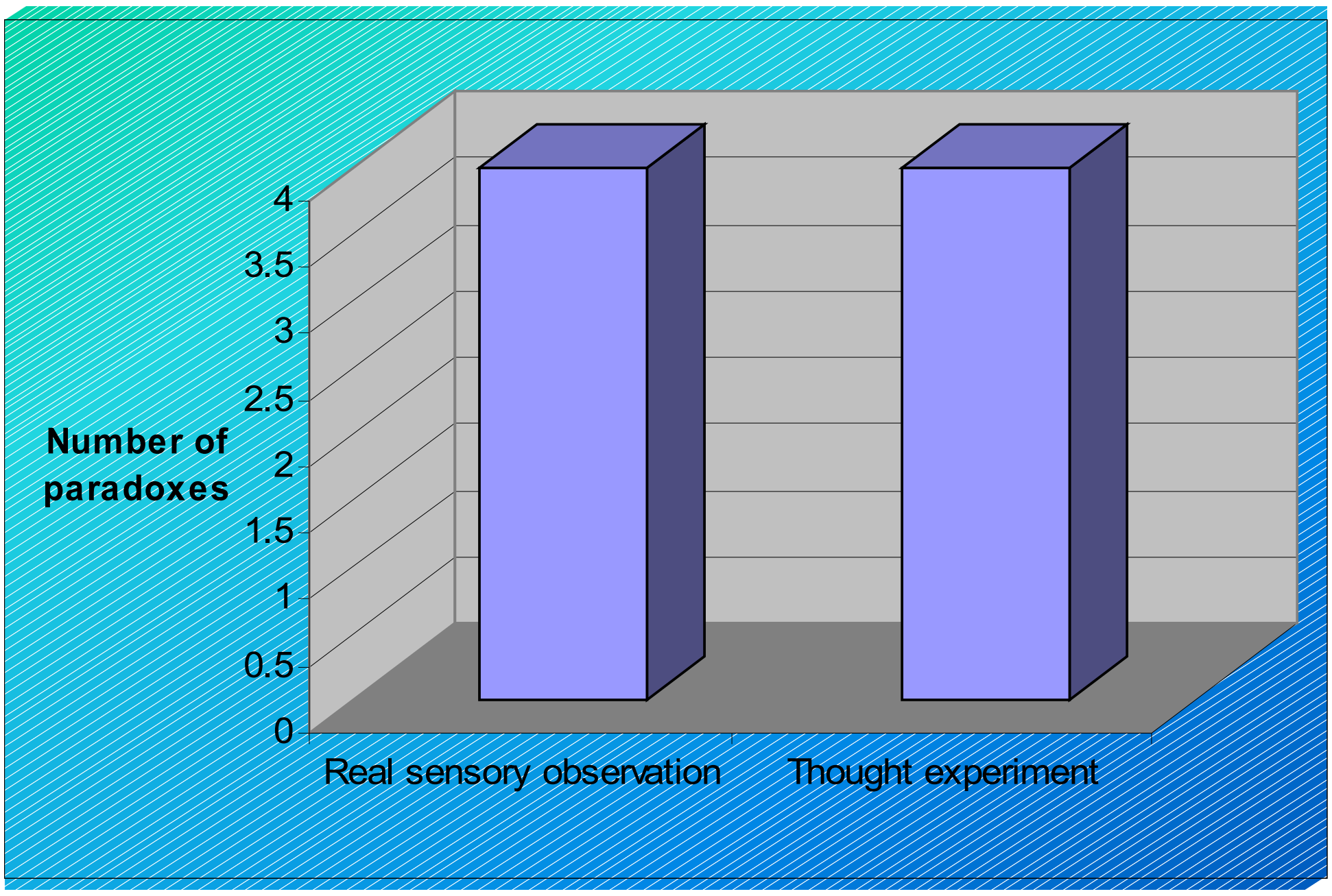

Graphic 3. Graphical representation of thought paradoxes in astrophysics.

# Tabelar view of Astrophysical paradoxes

| | GZK paradox | Seeliger's paradox | Olbers' paradox | Wheeler's paradox | Black hole information paradox | Eddington paradox | The faint young Sun paradox | Clausius' paradox |
|---|---|---|---|---|---|---|---|---|
| **type of paradoxes** | Paradox of assumption | Paradox of paradigm | Paradox of paradigm | Paradox of paradigm | Hierarchical paradox | Hierarchical paradox | Pseudo paradox | Paradox of paradigm |
| **solvability of paradoxes** | unsolved | exparadox | exparadox | exparadox | exparadox | exparadox | exparadox | exparadox |
| **thougth of paradoxes** | opservation | opservation | opservation | thought | thought | thought | thought | opservation |
| **experimental of paradoxes** | experimental | theory | theory | theory | theory | theory | theory | theory |